\documentclass[apj,numberedappendix]{emulateapj}

\usepackage{graphicx}
\usepackage{amsmath}
\usepackage{amssymb}

\newcommand{\Lpeak}{\ensuremath{L_{\rm peak}}}
\newcommand{\Le}{\ensuremath{L_{\rm e}}}

\newcommand{\Pt}{\ensuremath{P_{10}}}
\newcommand{\kapes}{\ensuremath{\kappa_{\rm es}}}

\newcommand{\vf}{\ensuremath{v_{\rm f}}}
\newcommand{\vs}{\ensuremath{v_{\rm s}}}
\newcommand{\vph}{\ensuremath{v_{\rm ph}}}

\newcommand{\Nifs}{\ensuremath{^{56}\mathrm{Ni}}}
\newcommand{\Cofs}{\ensuremath{^{56}\mathrm{Co}}}

\newcommand{\Esn}{\ensuremath{E_{\rm sn}}}
\newcommand{\Eint}{\ensuremath{E_{\rm int}}}

\newcommand{\Mej}{\ensuremath{M_{\rm ej}}}

\newcommand{\Msh}{\ensuremath{M_{\rm sh}}}

\newcommand{\Msun}{\ensuremath{M_{\odot}}}

\newcommand{\Lp}{\ensuremath{L_{\rm p}}}
\newcommand{\Ep}{\ensuremath{E_{\rm p}}}

\newcommand{\tp}{\ensuremath{t_{\rm p}}}

\newcommand{\td}{\ensuremath{t_{\rm d}}}
\newcommand{\te}{\ensuremath{t_{\rm e}}}
\newcommand{\tmax}{\ensuremath{t_{\rm peak}}}

\newcommand{\Rns}{\ensuremath{R_{\rm ns}}}
\newcommand{\Ons}{\ensuremath{\Omega_{\rm i}}}
\newcommand{\Bf}{\ensuremath{B_{14}}}
\newcommand{\vt}{\ensuremath{v_{\rm t}}}

\newcommand{\vsh}{\ensuremath{v_{\rm sh}}}

\newcommand{\be}{\begin{eqnarray}}
\newcommand{\ee}{\end{eqnarray}}

\begin{document}


\title{Supernova Light Curves powered by Young Magnetars }
\author{Daniel Kasen\altaffilmark{1,2} and  Lars Bildsten\altaffilmark{3}}
\bibliographystyle{apj}
\shorttitle{Magnetar Powered Supernovae}
\shortauthors{Kasen \& Bildsten}

\altaffiltext{1}
{Department of Astronomy and Astrophysics, University of California, Santa
  Cruz, CA }
  \altaffiltext{2}
{Hubble Fellow}
\altaffiltext{3}
{Kavli Institute for Theoretical Physics and Department of Physics, Kohn Hall, University of California, Santa
  Barbara, CA 93110}


\begin{abstract} 
We show that energy deposited into an expanding supernova remnant by a highly magnetic ($B\sim 5\times 10^{14}~{\rm G} $)  neutron star spinning at an initial period of $P_i\approx 2-20 \ {\rm ms} $ can substantially brighten the light curve.   For magnetars with parameters in  this range, the rotational energy is released on a timescale of days to weeks, which is comparable to 
 the effective diffusion time through the supernova remnant.   The late time energy injection can then be radiated without suffering overwhelming adiabatic expansion losses.  The magnetar input also produces a central bubble which sweeps  ejecta into an internal dense shell, resulting in a prolonged period of nearly constant photospheric velocity  in the observed spectra.  We derive analytic expressions for the light curve rise time and peak luminosity as a function of $B$,  $P_i$ and the properties of the supernova ejecta that allow for direct inferences about the underlying magnetar in bright supernovae.   We  perform numerical radiation hydrodynamical calculations of a few specific instances and compare the resulting light curves to observed events.   Magnetar activity is likely to impact more than a few percent of all core collapse supernovae, and may naturally explain some of the brightest events ever seen (e.g., SN~2005ap and SN~2008es) at $L \ga10^{44}$~ergs~s$^{-1}$.
 \end{abstract}

\keywords{
radiative transfer -- stars: neutron -- supernovae: general -- supernovae: individual (SN 2005ap, SN~2008es, SN~2007bi) 
} 


\section{Introduction} 
\label{sec:intro } 

Studies of soft gamma-ray repeaters and anomalous X-ray pulsars
reveal that $\sim10\%$ of newly born neutron stars  \citep{Kouv_98}
have dipole magnetic fields as high as  $B\sim 10^{14}-10^{15}~{\rm G}$
 for more than 1000 years after their birth \citep[see][]{Woods_06}. These ``magnetars'' rotate at
periods of $P=5-12\ {\rm s}$ at an age of  $1000-10,000 $ years.  Such highly magnetized neutron stars (NSs) were
theoretically predicted \citep{Duncan_92, Thompson_93}, and most of their activity (both sporadic and persistent) must be 
powered by the decay of these large magnetic fields.

What remains unknown is just how highly magnetized and rapidly rotating
these magnetars may be at ``birth''. Many \citep[see][]{Bodenheimer_74, Wheeler_00, Thompson_04} 
have investigated the possible impact on the central engine when the magnetar is so rapidly rotating (1-3 ms)  
and magnetized that its subsequent spin-down can power
the explosion. Cases this extreme may also be sources for ultra-high 
energy cosmic rays \citep{Arons_03} or deposit enough energy in the
 collapsing stellar envelope to favorably shape the deep interior \citep{Uzdensky_07, Bucci_09}
  for the production of a collimated relativistic flow needed for gamma-ray
bursts. Such events depend on the combination of rapid rotation and
high $B$ to achieve a measurable effect during the few seconds critical to the core collapse
mechanism. 

Building on the work of \cite{Gaffet_77a, Gaffet_77b}, we have found that  weaker magnetic
fields and less extreme spins (that do not alter the explosion
mechanism)  can dramatically impact  supernovae light curves, 
competing with the decay of radioactive $^{56}$Ni and thermal
energy in the expanding envelope. \cite{Maeda_07} previously  raised this possibility for the Type Ib SN~2005bf,
and \cite{Woosley_09} has independently shown their relevance as well.

 We show in \S \ref{sec:simple} that  when the timescale of the magnetar spindown, \tp, is comparable to the
 effective radiative diffusion time, \td, the resulting peak luminosity is
$\Lpeak \sim \Ep \tp /\td^2$, where \Ep\ is the magnetar rotational energy.
Magnetars with $10^{13}~{\rm G} < B<10^{16} \ {\rm G}$ and $P_i=1-30 \ {\rm ms}$  can produce $L_{\rm peak}>10^{42} \ {\rm erg \ s^{-1}}$.
We discuss the dynamics of the energy injection in \S \ref{sec:cavity} and show that the magnetar blows a central bubble in the SN ejecta, forming a dense inner shell of swept-up material which affects the spectroscopic evolution.  In \S \ref{sec:lightcurves}, we 
derive analytic expressions for the luminosity, $\Lpeak$,  and duration, $t_{\rm peak}$,  of magnetar powered supernovae.
We confirm these formulae with numerical radiation-hydrodynamical  calculations, and show how they can be inverted to infer $B$ and $P_i$ from a given light curve. We close in 
\S \ref{sec:conclusions} by discussing observed 
core-collapse SNe that may be powered this way, especially the ultra-bright SN~2005ap \citep{Quimby_07} and SN~2008es 
\citep{Gezari_09, Miller_09}.

\section{Magnetar Heating: Simple Estimates}
\label{sec:simple} 

In the simplest model, the core collapse mechanism has ejected an envelope
of mass $\Mej$ at a velocity \vt\  from a star of initial radius 
$R_0$. Within a few expansion times, $\te\sim R_o/\vt$, this 
ejecta will be undergoing self-similar adiabatic expansion, with an internal energy
$\Eint \sim \Esn (R_o/R)$, where  $\Esn\sim \Mej \vt^2/2$ and $R\sim v_t t $ is the remnant  size.   In the absence of magnetar (or $^{56}$Ni) heating, adiabatic expansion continues until the 
remnant is as old as the effective diffusion time $t_d\sim (\kappa \Mej/v_t c)^{1/2}$, where $\kappa$ is the opacity, after which the entropy is lost. Such thermally powered light-curves (e.g. Type IIp's) have a luminosity $L_{\rm th}\sim  \Esn \te/ \td^2$. The large amount of adiabatic expansion that 
has occurred by the time $t\sim t_d$ leads to low luminosities.

Now consider the impact of late time ($t \gg \te$)  energy injection from a young magnetar
with radius $\Rns=10 \ {\rm km}$ 
and initial spin $\Omega_i=2\pi/P_i$. 
The magnetar rotational energy is
\begin{equation}
\label{eq:ep} 
\Ep ={I_{\rm ns} \Ons^2\over 2} =   2 \times 10^{50} \Pt^{-2} ~\rm{ergs},
\end{equation}
where $\Pt = P_i/10$~ms and we set the NS moment of inertia to be 
$I_{\rm ns} = 10^{45}\ {\rm g \ cm^{2}} $. 
This magnetar  loses rotational energy at the rate set by magnetic dipole radiation (with the angle, $\alpha$,  between rotation and magnetic dipole fixed at $\sin^2\alpha=1/2$),  
injecting most of the energy into the expanding remnant on the spin-down timescale
\begin{equation} 
\label{eq:tp}
\tp = \frac{6 I_{\rm ns} c^3}{B^2 R_{\rm ns}^6 \Ons^2 }
= 1.3  \Bf^{-2} \Pt^2~  {\rm yr},
\end{equation}
where $\Bf=B/10^{14} \ {\rm G}$. 
To input this energy at a time $\tp \la \td$ requires a minimum $B$ field of 
\begin{equation} 
\label{eq:condition1} 
B > 1.8 \times 10^{14}~ \Pt ~\kapes^{-1/4}~M_5^{-3/8} E_{51}^{1/8}~{\rm G},
\end{equation} 
where  $\kapes = \kappa/0.2 \ {\rm cm^2 \ g^{-1}}$, $M_5 = \Mej/5~\Msun$ and $E_{51} = \Esn/10^{51}~{\rm ergs}^{-1}$. 
The required fields are in  the magnetar range.
 This late time entropy injection resets the interior energy 
scale to $\Eint \sim \Ep$ and overwhelms the initial thermal energy when  $\Ep>\Esn (t_e/t_p)$.  Thus even low magnetar energies $E_p<\Esn $  
play an important role. The resulting peak luminosity is
 \begin{equation} 
 \label{eq:lpeak} 
L_{\rm peak}\sim 
\frac{\Ep \tp}{\td^2} 
\sim 5 \times 10^{43}  \Bf^{-2} \kapes^{-1}  M_5^{-3/2} E_{51}^{1/2} {\rm erg \ s^{-1}},
 \end{equation} 
 which is primarily a function of the magnetic field.
This  shows that $L_{\rm peak}\sim 10^{43}-10^{45} \ {\rm erg \ s^{-1}}$ SNe can be achieved from magnetars with 
$\Bf=1-10$ and initial spins in the $P_i= 2-20$~ms range.

\section{Hydrodynamical Impact} 
\label{sec:cavity}

\begin{figure}
\includegraphics[width=3.5in]{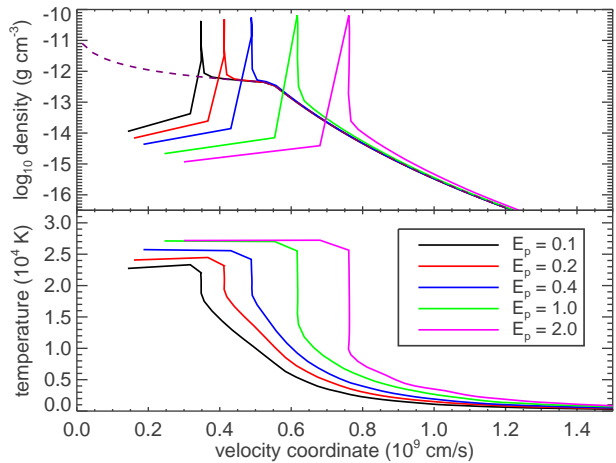}
\caption{Radiation-hydrodynamical calculations of the density (top) and temperature (bottom) of magnetar energized supernovae, one month after the explosion.  The supernova had $\Mej = 5~\Msun$ and $\Esn = 10^{51}$~ergs.   The magnetar had $\tp = 10^5$~sec and various values of \Ep, labeled in units of $10^{51}$~ergs.    The
dashed line in the top panel shows the unperturbed density structure, taken from Equation~(\ref{Eq:dens}). }
\label{Fig:ejecta}
\end{figure}

\begin{figure}
\includegraphics[width=3.5in]{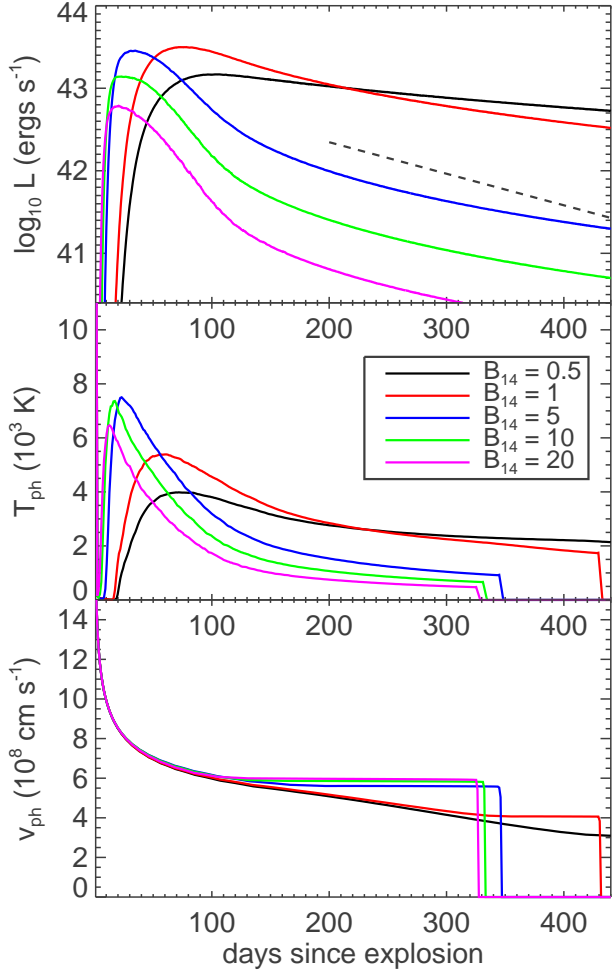}
\caption{Radiation hydrodynamical calculations of magnetar energized supernovae with
$\Mej = 5~\Msun$, $\Esn = 10^{51}$~ergs, and a density structure given by equation (\ref{Eq:dens}).  The magnetar
had $P_i = 5$~ms, and various magnetic field strengths.  {\it Top panel:}  Bolometric light curves.  The 
dashed line shows, for comparison, the energy deposition from $1~\Msun$ of \Nifs.  {\it Middle panel:}  Effective temperature.  {\it Bottom panel:} Velocity of the $e^-$ scattering photosphere at $\tau = 1$.  
}
\label{Fig:LCs}
\end{figure}

Our simple estimate  ignores the details of how the deposited energy is distributed throughout the interior of the expanding SNe remnant. Since the dissipation mechanism for the pulsar wind in this medium is poorly understood, we
assume the injected magnetar energy is  thermalized spherically at the base of the supernova ejecta. The remnant is assumed to be in homologous expansion with a shallow power law density
structure in the interior
\begin{equation}
\rho_0(v,t) =
\biggl[\frac{3 - \delta}{4 \pi}  \biggr]
\frac{\Mej}{v_t^3 t^3}  \biggl(\frac{v}{  v_t } \biggr)^{-\delta},
\label{Eq:dens}
\end{equation}
where $\vt = (2 \Esn/ \Mej)^{1/2}$ is the characteristic ejecta velocity, and the density
falls off sharply above \vt.

The central overpressure caused by the energy deposition blows a bubble in the SN remnant,
similar to the dynamics studied in the context of pulsar wind nebulae
\citep[e.g.,][]{Chevalier_77, Chevalier_92}.  As
this bubble expands, it sweeps up ejecta into a thin shell near the leading shock,
leaving the hot, low density  interior evident in the  1-D radiation hydrodynamical calculations of 
 Figure~\ref{Fig:ejecta}. 
In multi-dimensional calculations of pulsar wind nebulae, Rayleigh-Taylor instabilities broaden the shell and mix
the swept-up material \citep{Jun_98, Blondin_01}.

The bubble expansion will freeze out in Lagrangian coordinates
when the leading shock velocity  becomes comparable to  the local
velocity of the expanding SN ejecta.
The postshock pressure is $P = 2\gamma \rho_0 \vs^2 /(1+\gamma) = (8/7) \rho_0 \vs^2$ for a strong shock, and  the pressure of the energized cavity is  $P \approx \Ep/3 V$, where $V$ is the volume, implying a
shock velocity $\vs^2 = 7 \Ep/ 32 \pi R^3 \rho_0$.
The shock becomes weak when  $\vs \approx R/t$, which
determines the final velocity coordinate of the dense shell
\begin{equation}
\begin{split}
\vsh \approx  \vt
\biggl[ \frac{7}{16 (3 - \delta)} \frac{\Ep}{\Esn} \biggr]^{1/(5-\delta)},
~{\rm for}~{\Ep \la \Esn}.
\end{split}
\end{equation}
The weak dependence on \Ep , $\vsh \propto \Ep^{1/4}$, for $\delta = 1$, places 
\vsh \ near  \vt. The total mass swept up in the shell is $\Msh = \Mej (\vt/\vsh)^{3-\delta}$.

The magnetar does not affect the dynamics of the outer layers of the SN ejecta unless
 $\Ep \ga \Esn$, in which case
 the bubble expands beyond  \vt\ and  accelerates more rapidly down the steep outer density gradient. 
Essentially all of the ejecta is then swept up into the shell at a final
shell velocity 
\begin{equation}
\vsh \approx \vt [1 + \Ep/\Esn]^{1/2}~~{\rm for}~{\Ep \ga \Esn}.
\label{Eq:vsh}
\end{equation}
Both estimates for \vsh\ assume no radiative losses. 

The presence  of a dense shell has consequences for the supernova spectra. 
Initially the photospheric velocity, \vph, as measured from the Doppler shift of absorption line minima, decreases with time as the outer layers of ejecta become transparent.  Once \vph\ has receded to the shell velocity, however, it will remain constant (Figure~\ref{Fig:LCs}, bottom panel).
The spectra will then be characterized by relatively narrow but blueshifted absorption features,
and the spectral evolution will be notably slow.
The shell becomes optically thin to electron scattering at a time
\begin{equation}
t_{\tau = 1} = 326~M_5~E_{51}^{-1/2}
\kapes^{1/2}
~{\rm days}.
\end{equation}
Recombination
may hasten this transition. The photospheric velocity drops suddenly to zero after $t_{\tau=1}$,
 however because the line opacity in the shell remains optically thick longer,
the spectral features may not display any dramatic change for some time after.

\section{Light Curves}

We now derive  analytic expressions for the peak 
luminosity of a magnetar powered SNe using a one-zone model for the whole remnant.
The internal energy, \Eint,  is
governed by the first law of thermodynamics
\label{sec:lightcurves}
\begin{equation}
\frac{ \partial \Eint}{\partial t} =
- P \frac{ \partial  V}{\partial t}  + \Lp(t) -  \Le(t),
\label{Eq:first_law}
\end{equation}
where \Lp\ is the magnetar luminosity and \Le\ the radiated luminosity.
We assume that the magnetar energy is thermalized throughout the remnant, and that radiation pressure dominates, $P = \Eint/3V$. When the volume  increases as $V \propto t^3$, equation (\ref{Eq:first_law})  becomes
\begin{equation}
\frac{1}{t} \frac{ \partial } {\partial t} [\Eint t]= \Lp(t) - \Le(t).
\label{Eq:diff_Eint}
\end{equation}
The radiated luminosity, \Le, is approximated from  the diffusion equation
\begin{equation}
{\Le \over 4 \pi R^2} =   \frac{c}{3 \kappa \rho} \frac{\partial \Eint/V}{\partial r}
\approx \frac{c}{ 3\kappa \rho}  \frac{\Eint/V}{R},
\end{equation}
and rewritten using $R = \vf  t$, defining the effective diffusion time, \td
\begin{equation}
\Le = \frac{\Eint t}{\td^2}~~{\rm where}~~\td = \biggl[ \frac{3}{4 \pi} \frac{ \Mej \kappa }{\vf c}  \biggr]^{1/2},
\label{Eq:Le}
\end{equation}
where we take $\vf = [(\Ep + \Esn)/2\Mej]^{1/2}$ as the final characteristic ejecta velocity.
For the simple case where the magnetar injects a constant luminosity $\Lp = \Ep/\tp$
over a time \tp, and then shuts off, we find
\begin{equation}
\begin{split}
\Le(t) &= \frac{\Ep}{\tp} [1 - e^{-t^2/2 \td^2} ]~~~~~~~~~~~~ t < \tp ,  \\
\Le(t) &= \frac{\Ep}{\tp} e^{-t^2/2 \td^2} [e^{\tp^2/2 \td^2}  - 1] ~~~~~t > \tp.
\end{split}
\end{equation}
 This light curve peaks at a time \tp, then declines on the characteristic time scale \td.
For $\tp \ll \td$,   $\Lpeak = \Ep \tp/2 \td^2$, similar to
the estimate in \S 2. When $\tp \gg \td$, we find  $\Lpeak = \Ep/\tp$.

More generally,  the energy input from the magnetar persists for $t> \tp$, and is given by the spin-down formula
\begin{equation}
\Lp(t) = \frac{\Ep}{\tp}\frac{l-1}{(1 + t/\tp)^l},
\end{equation}
where $l=2$ for magnetic dipole spin down.   The energy input at late times may not be dynamically important, but it enhances the luminosity by continually heating the ejecta in a manner similar to the decay of  \Nifs.
No simple analytic solution for the light
curve exists for the general form of $\Lp(t)$,  but since radiative losses are minimal for times $t < \td$ we can derive approximate relations 
 by solving equation~(\ref{Eq:diff_Eint}) for the case
$\Le = 0$. The resulting internal energy can be evaluated at time \td\ in equation (\ref{Eq:Le})  to estimate the peak luminosity
\begin{equation}
\label{eq:magicform}
\begin{split}
\Lpeak &\approx  f \frac{\Ep \tp}{\td^2}
\biggl[\ln \biggl(1 + \frac{\td}{\tp} \biggr) - \frac{\td }{\td + \tp}  \biggr] ~~~~~~~~{l = 2},\\
\Lpeak &\approx  f \frac{\Ep \tp}{\td^2} \frac{1}{l - 2} \biggl[1 - \frac{\td/\tp(l-1) + 1}{(1 + \td/\tp)^{l-1}}
\biggr]~~~~{l > 2},
\end{split}
\end{equation}
where the correction factor $f$ will be calibrated by comparison to numerical simulation.
  In general, \Lpeak\ decreases
as $l$ increases, as more of the energy is deposited at earlier times and suffers greater adiabatic losses. 
 
 \begin{figure}[t]
\includegraphics[width=3.5in]{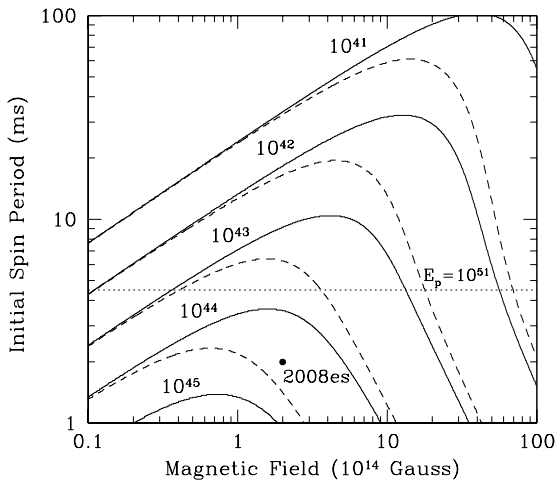}
\caption{Required B and $P_i$ needed  to achieve a given $\Lpeak$. The lines are contours of constant $\Lpeak$  assuming $\Esn=10^{51}~{ \rm ergs}$ and $\Mej=5~M_\odot$ (solid) or $\Mej=20~M_\odot$ (dashed) from equation (\ref{eq:magicform}).  Regions to the right of the knee have $t_p<t_d$, whereas regions to the left of the knee  have  $t_p>t_d$. The horizontal dotted line shows where $E_p=10^{51}~{\rm ergs}$. \label{Fig:lum} }
\end{figure}

At the peak of the light curve, the radiated luminosity equals the instantaneous magnetar luminosity, $\Lpeak = \Lp(\tmax)$,  the  general expression of ``Arnett's law"  (Arnett 1979).    This follows  from equation (\ref{Eq:diff_Eint}), since equation (\ref{Eq:Le}) implies that the maximum of \Le\ occurs when $\partial (\Eint t) /\partial t = 0$, yielding the time of maximum in the light curve
\begin{equation}
\tmax = \tp \biggr( \biggl[ \frac{(l-1)\Ep}{\Lpeak \tp} \biggr]^{1/l} - 1 \biggr).
\label{eq:tpeak}
\end{equation}
For $\tp \ll \td$ the light curve peaks at  $\tmax \approx \td f^{-1/2}[\ln(\td/\tp)-1]^{-1/2}$ (assuming $l=2$), whereas for
 $\tp \gg \td$ the peak occurs at $\tmax \approx \tp   (\sqrt{2/f}-1)$. 
 
Figure~\ref{Fig:LCs} shows 1-D radiation hydrodynamical calculations 
for  $\Mej = 5~\Msun, \Esn = 10^{51}$~erg,
and central magnetars ($l=2$) with $P_i = 5$~ms. A grey opacity
 $\kappa = 0.2 \ {\rm cm^2 \ g^{-1}} $ was assumed. 
 The simple one zone model works remarkably well at predicting $\Lpeak$ and $\tmax$ and
comparison with the numerical models fixes the value of  $f = (l + 1)/2$.
At late times ($t > t_{\tau =1}$) when the SN becomes optically thin, the light curve tracks the magnetar luminosity, $L \sim t^{-2}$, which is similar to the curve of \Cofs\ decay.  Late time measurements of the bolometric light curve could discriminate the two energy sources, though it is not clear that the assumptions of complete thermalization and constant $l=2$ spindown will hold at these late times. 
 
 In Figure~\ref{Fig:lum}, we use equation (\ref{eq:magicform}) to find the locus in the $P_i-B$ space (assuming $l=2$) needed to reach a certain $\Lpeak$
 in a supernova with $\Esn=10^{51}$~ergs and  $\Mej=5~\Msun$ or  $\Mej=20~\Msun$.  A larger \Mej\ increases \td, which reduces \Lpeak\ for a given set of magnetar parameters. Magnetars with $P_i \la 5$~ms (below the dotted line)  dump enough  energy to increase the ejecta velocity, shortening  \td. 
The lines merge for low $B$ as they asymptote to $L_{\rm peak}\rightarrow L_p$.  

We can also invert the problem and use the measured values of 
\Lpeak\ and \tmax\  for an individual supernova 
to infer $B$ and $P_i$. Figure~\ref{Fig:waycool} uses equations (\ref{eq:magicform}) and (\ref{eq:tpeak}) to illustrate how $\Lpeak$ and $\tmax$ vary with $B$ and $P_i$. This ``mapping'' allows for an assessment to be made of the magnetar's properties and illuminates which numerical calculations should be done. We placed the observed values for 2008es on this plot, motivating the numerical results we show in the following section. A different plot would need to be made for different $\Mej$ and $\Esn$. 

\section{Discussion and Conclusion}
\label{sec:conclusions}

\begin{figure}
\includegraphics[width=3.5in]{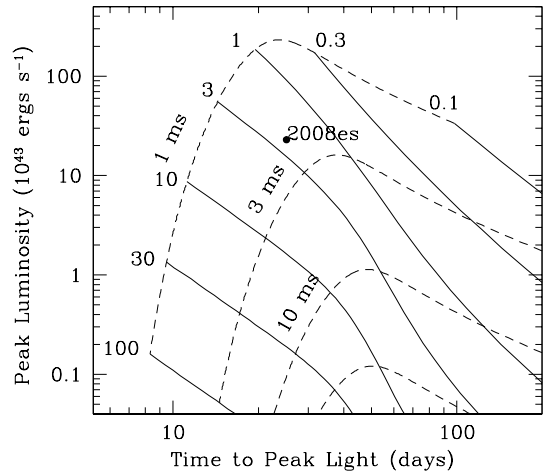}
\caption{The dependence of $\Lpeak$ and $\tmax$ on the initial magnetar spin and B field. The solid lines are for fixed $B_{14}=100,30,10,3,1, 0.3$ and 0.1 and varying spin period, whereas the dashed lines are for a fixed $P_i=1,3,10$ and 30 ms  and varying $B$.  This calculation assumed $\Esn=10^{51}~{ \rm ergs}$ and $\Mej=5~M_\odot$.  \label{Fig:waycool} }
\end{figure}

We have shown that rotational energy deposition from magnetar spin-down  with initial spin periods $<30 \ {\rm ms}$  can substantially modify the thermal evolution of an expanding SNe remnant. For magnetars in this range, the peak luminosity reaches $10^{42} -10^{45}~{\rm erg \ s^{-1}}$  ($M_{\rm Bol}= -16.3$ to $-23.8$), substantially impacting the typical core-collapse SNe lightcurve, whether it is a Type II or a Ibc event.    The highest luminosities occur when $\tp \sim \td$, in which case the total energy radiated in the light curve is $E_{\rm rad} \sim \Lpeak \td \sim \Ep/3$.  The maximal spin of  a NS is  around 1 ms, so $E_{\rm rad}$ cannot exceed $\sim 10^{52}$~ergs; supernova radiating larger energies can not be explained by this mechanism.
 Though we know that $\sim 10\%$ of core collapse events make magnetars, we do not know the distribution of initial spin periods, so the prevalence of light curve dominance is difficult to predict.

\begin{figure}
\includegraphics[width=3.5in]{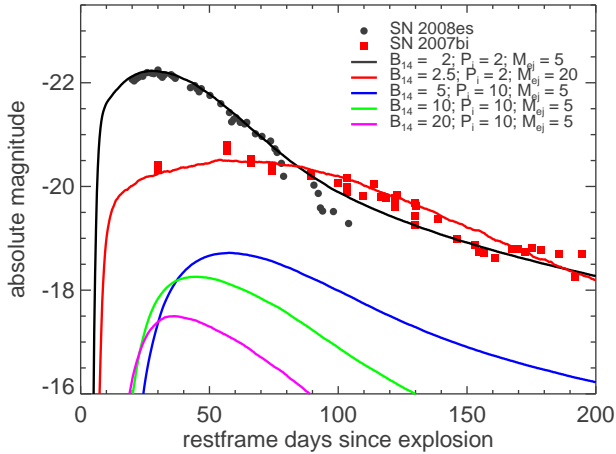}
\caption{Bolometric light curve calculations of magnetar energized supernovae compared to observed events.
A constant opacity $\kappa = 0.2$~g~cm$^{-2}$ is assumed. 
Black circles show V-band observations of the luminous Type~IIL~SN2008es \citep{Gezari_09} with an assumed rise time of 25~days.  Red squares show R-band observations of the Type~Ic SN~2007bi \citep{Gal-Yam_09} with an assumed rise time of 50 days.   }
\label{Fig:compare}
\end{figure}

For stars with remaining hydrogen, magnetar injection may explain the brighter ($M_B \sim -19$) subclass of Type II-L SNe  noted by \cite{Richardson_02}, i.e. 1961F, 1979C, 1980K, 1985L. The light curves of these events are difficult to explain in standard explosion models unless extreme progenitor radii ($R > 2000~R_\odot$)  are assumed \citep{Blinnikov_93}.  Figure~\ref{Fig:compare} shows that a magnetar with relatively modest rotation, $P_i = 10 $~ms, in a $\Mej = 5~\Msun$ supernova can reach similar luminosities.  Events brighter than $M_{\rm Bol}=-21$ ($L>8\times 10^{43} ~{\rm erg \ s^{-1}}$), such as the ultrabright Type II-L SN~2005ap \citep{Quimby_07} and SN~2008es \citep{Gezari_09, Miller_09} require an initial magnetar spin of $<5 \ {\rm ms}$. Motivated by Figure~\ref{Fig:waycool}, we 
found an excellent fit to the SN~2008es light curve with  $\Bf = 2 $, $P_i = 2$~ms and $\Mej=5 \Msun$. 
Such rapidly rotating magnetars must be rare, as \cite{Vink_06} found that the galactic supernovae remnants of known magnetars were explained with typical explosion energies of $10^{51} \ {\rm ergs}$. This rarity is consistent with the
specific volume rate of these events; current estimates put them at no more than $\sim 1\%$ \citep{Miller_09, Quimby_09} of  the local core collapse rate.

Debate remains \citep[see][]{Klose_04, Gaensler_05,Davies_09}  as to whether magnetars are preferentially formed from the most massive stars that collapse to NSs. If so, then we might see a prevalence of magnetar dominated light curves amongst the Ib/c SNe, which may partially explain the wide light curve diversity noted in this class. Some extreme SN~Ic, such as SN2007bi \citep{Gal-Yam_09, Young_09} and SN~1999as \citep{Knop_99}, which remained very bright for a long time, have been  claimed to be the pair instability explosion of a $\sim 100~\Msun$ star producing nearly  $5~\Msun$ of \Nifs.  Figure~\ref{Fig:compare} shows that the light curve could alternatively be explained for a supernova with  $\Mej = 20$~\Msun\ forming a magnetar with $\Bf = 2, P_i = 2.5$~ms.   The spectra of SN~1999as also revealed a slowly evolving photospheric velocity and  narrow, blueshifted absorption features, suggestive of a dense shell like that predicted here \citep{Kasen_thesis}.  On the other hand, the magnetar model may have trouble reproducing the strong iron emission lines seen in the nebular phase spectrum of SN~2007bi.

Our initial investigations have revealed that if an appreciable fraction of highly magnetic NSs are born rapidly rotating, then we should find evidence for them in the plethora of supernovae surveys, such as the Palomar Transient Factory  \citep{Law_09}. Many open questions remain on the theoretical side, especially how the outgoing pulsar wind thermalizes in the remnant, whether there are substantial Rayleigh-Taylor instabilities, and how these could manifest themselves in the observed spectra both at late times and during the photospheric phase. Our work has outlined the regimes of relevance, and will guide future large scale computations through parameter space in an informed manner.

\acknowledgments
We thank David Kaplan, Eliot Quataert, and Stan Woosley for helpful discussions.
Support for D.K was provided by NASA through Hubble fellowship grant
\#HST-HF-01208.01-A awarded by the Space Telescope Science Institute,
which is operated by the Association of Universities for Research in
Astronomy, Inc., for NASA, under contract NAS 5-26555. 
This research has been supported by the DOE
SciDAC Program (DE-FC02-06ER41438) and by the National Science Foundation under
grants PHY 05-51164 and AST 07-07633.

\end{document}